\documentclass[fleqn,twoside]{article}
\usepackage{espcrc2}

\usepackage{graphicx}
\usepackage[figuresright]{rotating}

\newcommand{\AmS}{{\protect\the\textfont2
  A\kern-.1667em\lower.5ex\hbox{M}\kern-.125emS}}

\title{X-Ray Sources Overdensity Around the 3C 295 Galaxy Cluster}

\author{V. D'Elia\address[MCSD]{INAF - Osservatorio Atronomico di Roma \\ 
        Via di Frascati, 33, I-00040, Monte Porzio Catone, Rome, Italy}%
        F. Fiore
        M. Elvis M. Cappi S. Mathur P. Mazzotta E. Falco}

\begin{document}

\begin{abstract}
We present a statistical analysis of the Chandra observation of 
the source field around the 3C 295 galaxy cluster ($z=0.46$).  
The logN-logS of this field is in good agreement with that computed
for the Chandra Deep Field South in this work and in previous ones. 
Nevertheless, the logN-logS computed separately for the four ACIS-I
chips reveals that there is a significant excess of sources to the
North-North East and a void to the South of the central cluster. Such
an asymmetric distribution is confirmed by the two-dimensional
Kolmogorov-Smirnov test, which excludes ($P \sim 3\%$) a uniform
distribution. In addition, a strong spatial correlation emerges from
the study of the angular correlation function of the field: the angular
correlation function is above that expected for X-ray sources on a few
arcmin scales. 
In synthesis, the present analysis may indicate a filament of the
large scale structure of the Universe toward 3C 295. This kind of
studies may open-up a new way to map (with high efficiency)
high-density peaks of large scale structures at high-z.

\vspace{1pc}
\end{abstract}

\maketitle

\begin{figure}[htb]
\includegraphics[scale=0.39]{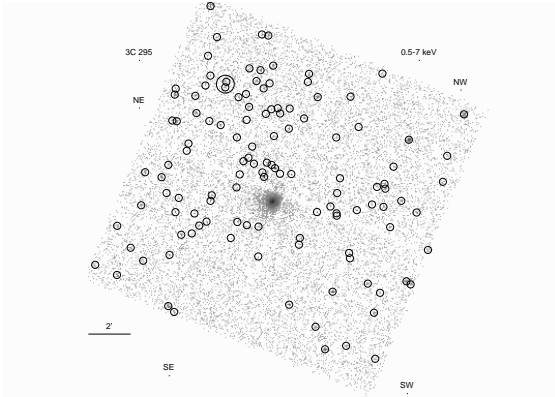}
\caption{The {\it Chandra} 3C 295 field in the $0.5-7$ keV band. 
Circles represent the sources detected with the
wavelet-based source detection code `PWdetect'. The largest circles in
the left and bottom panels indicates a very extended source, possibly
a group of galaxies.  The brightest source in the center of the field
is the cluster of galaxies 3C 295.} 
\label{fig:largenenough}
\end{figure}

\section{Introduction}

N-body and hydrodynamical simulations show that high redshift
clusters of galaxies lie at the nexus of several filaments of galaxies
(see e.g. Peacock 1999). Such filaments map out the ``cosmic web'' 
of voids and filaments of the large scale structure of the Universe. Thus, rich
clusters represent good indicators of regions of sky where several
filaments converge. The filaments themselves could be mapped out by
Active Galactic Nuclei (AGNs), assuming that AGNs trace galaxies. 

Cappi et al. (2001) studied the distribution of the X-ray sources
around 3C 295 ($z=0.46$). The clusters was observed with the ACIS-S CCD
array for a short exposure time ($18$ ks). 
They  found a high source surface density in the 0.5-2 keV band which
exceeds the ROSAT (Hasinger et al. 1998) and Chandra (Giacconi et
al. 2002) logN-logS by a factor of $2$,  with a significance of $2\sigma$ . 

In this work an analysis of a deeper  ($92$ ks) Chandra observation of
3C 295 is performed, to check whether the overdensity of such a field
in the $0.5-2$ keV band is real or not, to define the structure of the
overdensity, and to extend the above considerations to the $2-10$ keV
band.

\begin{figure}[htb]
\includegraphics[scale=0.39]{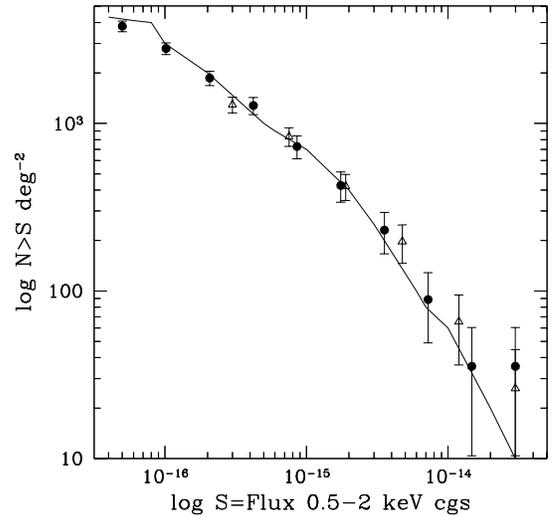}
\caption{The 3C 295 (open triangles) and {\it Chandra} Deep Field
South (filled circles) logN-logS in the $0.5-2$ keV band. Errors
represent $1\sigma$ confidence limit. Solid lines represent the
CDFS LogN-LogS from Rosati et al. (2002).}
\label{fig:largenenough}
\end{figure}

\begin{figure}[htb]
\includegraphics[scale=0.39]{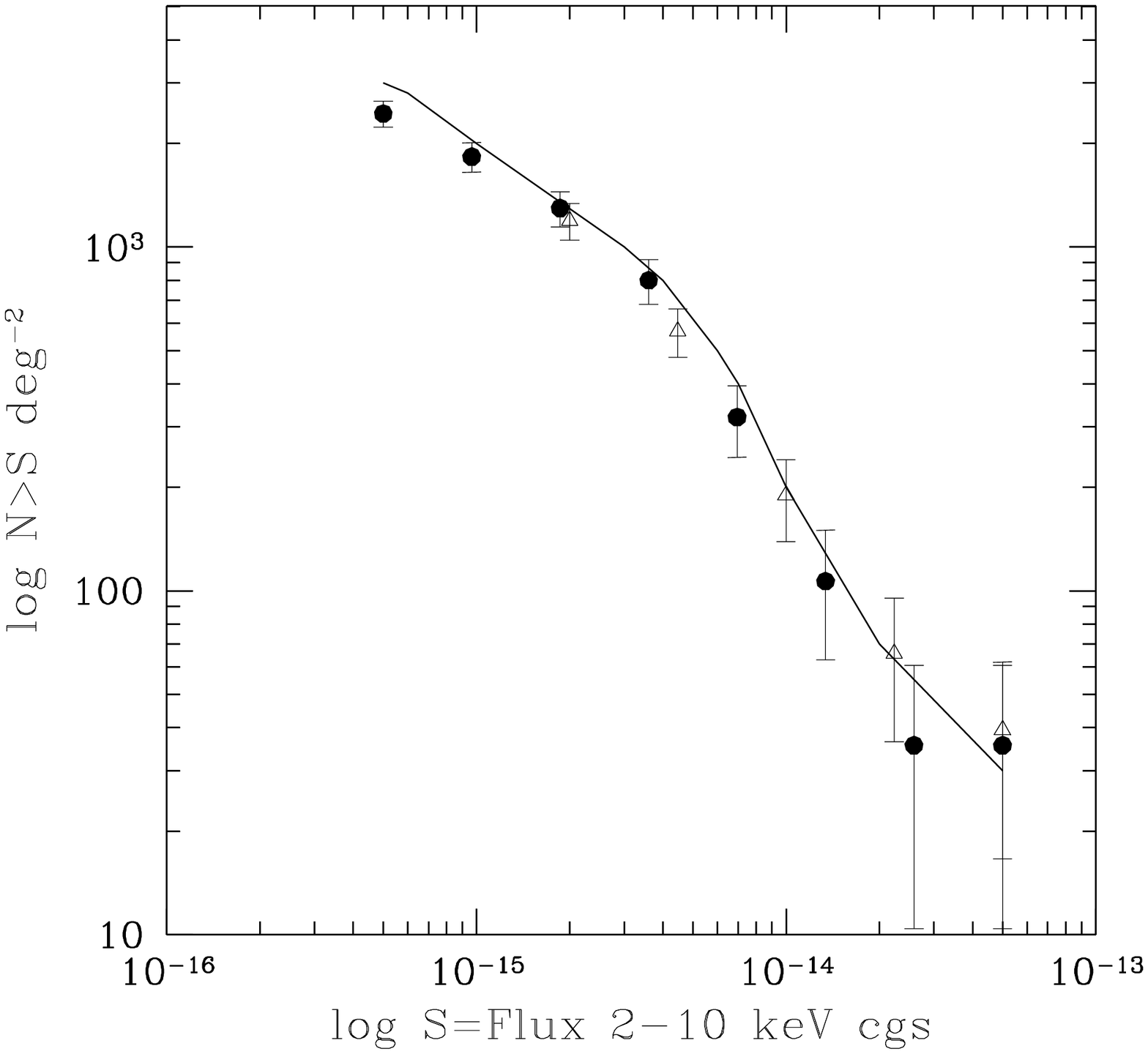}
\caption{The 3C 295 (open triangles) and {\it Chandra} Deep Field
South (filled circles) logN-logS in the $2-10$ keV band. Errors
represent $1\sigma$ confidence limit. Solid lines represent the
CDFS LogN-LogS from Rosati et al. (2002).}
\label{fig:largenenough}
\end{figure}

\section{Observation and Data reduction}

Chandra observed the $16' \times 16'$ field around the 
3C 295 cluster with ACIS-I on May 18, 2001, for $92$ ks. 

The data reduction was carried out using the Chandra Interactive
Analysis of Observations software version 2.1.3. 

All the data analysis has been performed separately in the $0.5-2$ keV,
in the $2-7$ keV and in the $0.5-7$ keV bands. 

An identical analysis was performed for the Chandra Deep Field South
(CDFS) in the $0.5-2$ keV and $2-7$ keV bands for comparison, and to check
our analysis methods. 

The source detection was carried out using the `PWDetect' algorithm
(Damiani et al. 1997). We identified $89$ sources
in the $0.5 -2$ keV band, $71$ sources in the $2 -7$ keV band and $121$
sources in the $0.5 -7$ keV band (see fig. 1). 

The counts in the $0.5-2$ keV, $2-7$ keV and $0.5-7$ keV
bands were converted in $0.5 -2$ keV, $2 -10$ keV and $0.5 -10$ keV fluxes
using conversion factors appropriate for a $\gamma = 1.8$ power law spectrum
with a galactic absorption toward the 3C 295 field of 
$N_H = 1.33 \times 10^{20}$ cm$^{-2}$. Such values take into account
the quantum efficiency degradation of the CCD.

\section{Analysis}

The following analysis has been perfomed: 

The sky coverage has been computed for 3C 295 and CDFS fields.

The LogN-LogS in the soft and hard bands for the whole filelds have
been produced (figs. 2 and 3).

The LogN-LogS in the soft, hard and whole bands have been produced
separately for each ACIS-I chip of the 3C 295 observation (see fig. 4
for the $0.5-10$ keV band).

3C 295 chip-to-chip and CDFS LogN-LogS have been fitted to evaluate
slopes and normalizations (fig. 5 for the $0.5-10$ keV band).

A two dimensional Kolmogorov-Smirnov test has been applied to check
whether the sources were uniformly distributed or not (tab. 1).

The angular correlation function has been computed in order to
estimate at which scales the sources in the two fields were clustered;
the errors associated to the function (fig. 6) have been calculated using both
poisson and bootstrap statistics (Barrow, Bhavsar \& Sonoda 1984).

\begin{figure}[htb]
\includegraphics[scale=0.39]{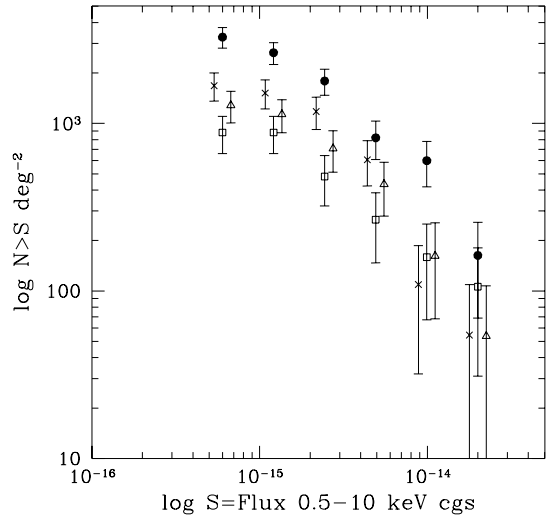}
\caption{ The mean (whole field) 3C 295 logN-logS in the 
$0.5-10$ keV band, calculated for each ACIS-I chip
separately. Filled circles represent counts for the NE chip, open
triangles for the NW, open squares for the SW and crosses for the SE.
For clarity reasons, the points for SE chip and the NW have been
shifted slightly to the left and to the right, respectively. Errors
represent $1\sigma$ confidence limit.}
\label{fig:largenenough}
\end{figure}

\begin{figure}[htb]
\includegraphics[scale=0.39]{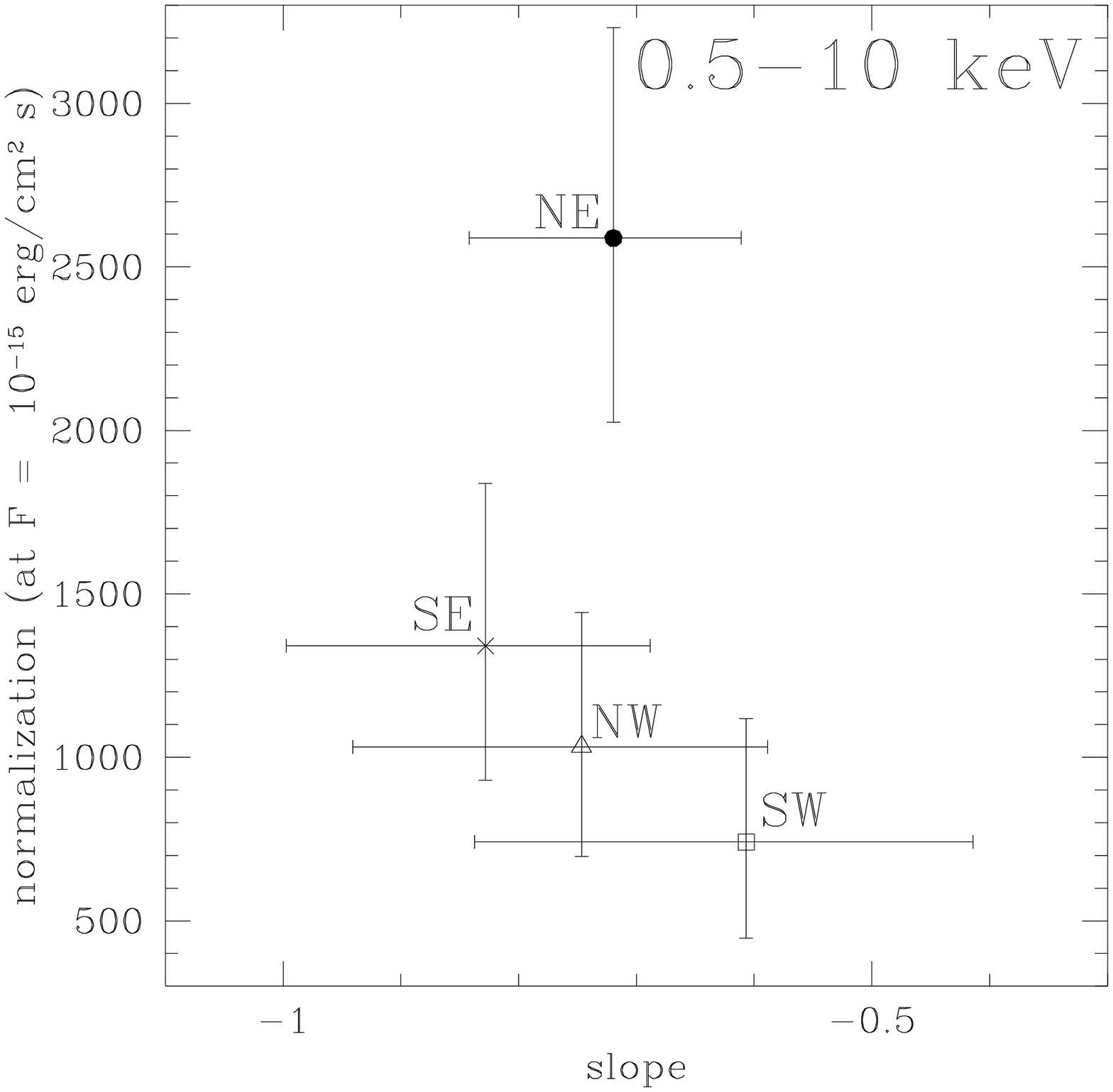}
\caption{Results of the power law fits to the four logN-logS chips in
the $0.5-10$ keV band. x axis plots the slope of the power law,
while on the y axis is plotted the normalization at 
$ 10^{-16}$ ergs cm$^{-2}$
s$^{-1}$.  Filled circles represent the NE chip, open
triangles the NW, open squares the SW and crosses the SE. 
Errors are the $90\%$ confidence limit.}
\label{fig:largenenough}
\end{figure}

\section{Main results}

The following results have been achieved: 

3C 295 and CDFS LogN-LogS are in very good agreement both in the soft
and hard band, and with the CDFS LogN-LogS by Rosati et al. 2002
(figs. 2 and 3).

The 3C 295 LogN-LogS in the soft, hard and broad bands computed
separately for each ACIS-I chip show an overdensity of sources in the
North-East (NE) chip (fig. 4 for the $0.5-10$ keV band) 
which reflects the clustering of sources
clearly visible in fig. 1.

The discrepancy between the normalization of the LogN-LogS for the NE
and SW chip is $3.2\sigma$, $3.3\sigma$ and $4.0\sigma$ in the soft, 
hard and broad band, respectively; fig. 5 shows the 
normalization vs. slope plot for the broad band. 

The two dimensional Kolmogorov-Smirnov test shows that there is a
considerable probability that CDFS sources are uniformly distributed
($P \sim 15\%$ in the soft and hard bands), while the probability that the
3C 295 sources are uniformly distributed is only a few per cent, and
drops below 1 if we consider the $0.5 - 7$ keV band (tab. 1).  

The angular correlation function of the 3C 295 sources features a strong
signal on scales of a few arcmins, and
also on lower scales in the $0.5 - 7$ keV band (fig. 6). Moreover, the
function is above the value found by Vikhlinin \& Forman (1995) for a
large sample of ROSAT sources. On the other hand, no signs of a
similar behavior is featured by the CDFS.

\begin{figure}[htb]
\includegraphics[scale=0.39]{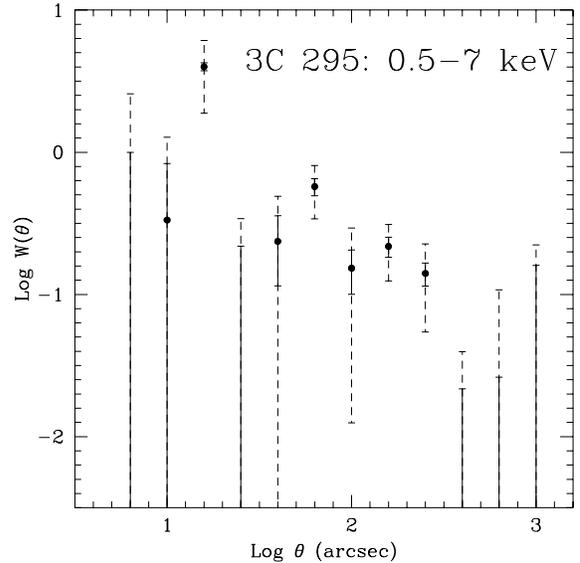}
\caption{The 3C 295 angular correlation function in the $0.5-7$ keV
 band. Solid error bars are Poisson;
dashed error bars are bootstrap.}
\label{fig:largenenough}
\end{figure}

\medskip
\medskip
\medskip

\begin{tabular}{|l|c|c|}

\hline
 & 3C 295 	& CDFS 	\\ 
\hline
$0.5-2$ keV	& $3.09\;10^{-2}$	& $0.13$\\
\hline
$2-7$ keV	& $3.84\;10^{-2}$	& $0.17$\\
\hline
$0.5-7$ keV	& $6.32\;10^{-3}$	& - \\
\hline
\end{tabular}

\medskip
{\bf Tab.1 KS test for CDFS and 3C 295 fields}

\section{Discussion}

In this work we studied in great quantitative details
the excess of sources clearly visible in the upper left corner of the
Chandra observation of the 3C 295 galaxy cluster field (see
fig. 1). Since N-body and hydrodynamical
simulations show that clusters of galaxies lie at the nexus of several
filament, this excess could represent a filament of the large scale
structure of the Universe.

Moreover, if the redshift of our sources were 
the same of 3C 295 ($z=0.46$), we could compute the galaxy overdensity
of the field assuming a spherical distribution. Under these
assumption, we obtain a density contrast of $\sim 70$ which, despite
large uncertainties, is 
intriguingly close to the expected
galaxy overdensity of filaments $\sim 10 \div 10^2$,
and much smaller than the 
overdensities of clusters of galaxies ($\sim 10^3 \div  10^4$).

Thus, in order to study this candidate filament, further Chandra 
observations and optical identifications are pursued to obtain the
redshift of the sources, to map out the 
3C 295 region, and delimiting the filament properties up to scales 
of $24$ arcmin (i.e., $\sim 6$ Mpc) from the 3C 295 cluster.

More details on the present work can be found in D'Elia et al., A\&A, 
submitted.


\end{document}